\documentstyle[epsfig,12pt]{article}

\newcommand{\be}{\begin{equation}}
\newcommand{\ee}{\end{equation}}
\newcommand{\bea}{\begin{eqnarray}}

\newcommand{\eea}{\end{eqnarray}}

\def\gappeq{\mathrel{\rlap {\raise.5ex\hbox{$>$}}
{\lower.5ex\hbox{$\sim$}}}}

\def\ket#1{
\left|{#1} \rangle \right . }

\def\lappeq{\mathrel{\rlap{\raise.5ex\hbox{$<$}}
{\lower.5ex\hbox{$\sim$}}}}

\begin{document}

\date{}                                 

\begin{titlepage}
\samepage
\setcounter{page}{1}

\rightline{ACT-02/02}
\rightline{CTP-TAMU-06/02}

\vspace{.05in} 

\begin{centering} 

{\Large \bf QED-Cavity model of microtubules implies dissipationless energy transfer and biological quantum teleportation }
\vspace{.20in}

\vspace{0.06in}        
{\bf Nick E. Mavromatos} \\

\vspace{.06in}
{\it Department of Physics, Theoretical Physics Group, \\
University of London, King's College \\
Strand, London WC2R 2LS, United Kingdom.\\
{\rm e-mail: Nikolaos.Mavromatos@kcl.ac.uk }\\} 

\vspace{0.06in} 
{\bf Andreas Mershin}\\
\vspace{.06in}
{\it Center for Theoretical Physics,\\
            Department of Physics, Texas A\&M University,\\
            College Station, TX 77843-4242, USA\\
{\rm e-mail: mershin@physics.tamu.edu}\\}

\vspace{0.06in} 
and \\
\vspace{0.06in} 
{\bf Dimitri V. Nanopoulos } \\      
\vspace{0.06in} 
{\it Center for Theoretical Physics,\\
            Department of Physics, Texas A\&M University,\\
            College Station, TX 77843-4242, USA\\}
{\it Astro Particle Physics Group,\\
            Houston Advanced Research Center (HARC),\\
            The Mitchell Campus,\\
            Woodlands, TX 77381, USA\\}

{\it  Academy of Athens, Chair of Theoretical Physics,\\ 
            Division of Natural Sciences,\\
            28 Panepistimiou Avenue, Athens 10679, Greece\\

{\rm e-mail: dimitri@physics.tamu.edu} \\}

\vspace{.38in}


\vspace{.24in}



\noindent {\bf Abstract} \\

\end{centering} 

{\small\em We refine 
a QED-cavity model of microtubules (MTs), proposed earlier by two 
of the authors (N.E.M. and D.V.N.),   
and suggest 
mechanisms for the formation of biomolecular mesoscopic coherent
and/or entangled quantum states, which may  
avoid decoherence for times comparable to biological characteristic times. 
This refined model predicts dissipationless
energy transfer along such ``shielded" macromolecules at near room
temperatures as well as quantum teleportation of states across MTs
and perhaps neurons.}

\vspace{.24in}


\end{titlepage}


\section{Introduction}

Observable quantum effects in biological matter such as proteins are naively
expected to be strongly suppressed, mainly due to the macroscopic
nature of most biological entities as well as the fact that such
systems live at near room temperature. These conditions normally
result in a very fast {\it collapse} of the pertinent
wave-functions to one of the allowed classical states. However,we suggest that
under certain circumstances it is in principle possible to obtain the necessary
isolation against thermal losses and other environmental
interactions, so that {\it meso- and macroscopic}
quantum-mechanical coherence, and conceivably entanglement
extending over scales that are considerably larger than the atomic
scale, may be achieved and maintained for times comparable to the
characteristic times for biomolecular and cellular processes.

In particular, we have shown ~\cite{mn} how microtubules (MTs)
~\cite{dustin} can be treated as quantum-mechanically isolated
(QED) cavities, exhibiting properties analogous to those of
electromagnetic cavities routinely used in quantum optics
~\cite{haroche,rabi,agar,rabiexp}. 
Recently, our speculative model has been
supported by some indirect experimental evidence. It has been
experimentally shown ~\cite{caes}, that it is possible to maintain
partial entanglement of the bulk spin of a macroscopic quantity of
Caesium (Cs) atoms $(N=10^{12})$, {\it at room temperature}, for a relatively
long time (0.5ms). Note that in this experiment, the large
quantity of atoms was of paramount importance in creating and
maintaining the entanglement, and even though the gas samples were
in constant contact with the environmental heat bath, by using a
careful experimental arrangement, Julsgaard et.al. managed to
detect the existence of entanglement for a much longer time than
one would intuitively expect. Here, we outline the
main features of our model of the quantum mechanical properties of
MTs (described in detail in ~\cite{mn}), and we exhibit the
relevance of the Julsgaard et. al. experiment to our model. A
direct consequence of our model for MTs as QED cavities is that
virtually every experimentally known QED-cavity-based observation
may have an analogue in living MTs and we show this analytically
with the specific case of intra- and inter-cellular dissipation-less energy transfer and quantum {\it teleportation} of coherent quantum states.

Energy transfer across cells, without dissipation, had been
first speculated to occur in biological matter by
Fr\"ohlich~\cite{Frohlich}. The phenomenon conjectured by
Fr\"ohlich was based on a one-dimensional superconductivity model:
a one dimensional electron system with holes, where the formation
of solitonic structures due to electron-hole pairing results in
the transfer of electric current without dissipation. Fr\"ohlich
suggested that, if appropriate solitonic
configurations are formed inside cells, energy in biological matter
could also be transferred without any dissipation (superefficiently). 
This idea has lead theorists to construct various models for cellular energy
transfer, based on the formation of kink classical
solutions~\cite{lal}.

In these early works no specific microscopic models had been
considered~\cite{lal}. In 1993 Sataric et. al. constructed a
{\it classical} physics model for microtubule dynamics \cite{mtmodel},
in which solitons transfer energy across MTs without dissipation.
In the past, we have considered the {\it quantum
aspects} of this one-dimensional model, and developed a framework
for the consistent quantization of the soliton solutions\cite{mn}. Our work
suggested that such semiclassical solutions may emerge as a result
of `decoherence' due to environmental interactions, in agreement
with ideas in ~\cite{zurek}.

The basic assumption used in creating the model of ref. \cite{mn}
was that the building blocks of MTs, the tubulin molecule
dimers, can be treated as elements of Ising spin chains (one-space-dimensional
structures). The interaction of each tubulin chain (protofilament) with the neighboring
chains and the surrounding water environment has been accounted for by
suitable potential terms in the one-dimensional Hamiltonian. The
model describing the dynamics of such one-dimensional
sub-structures was the ferroelectric distortive spin chain model
of ref. \cite{mtmodel}.

\begin{figure}[!ht]
\begin{centering}
\scalebox{0.6}{\epsfig{figure=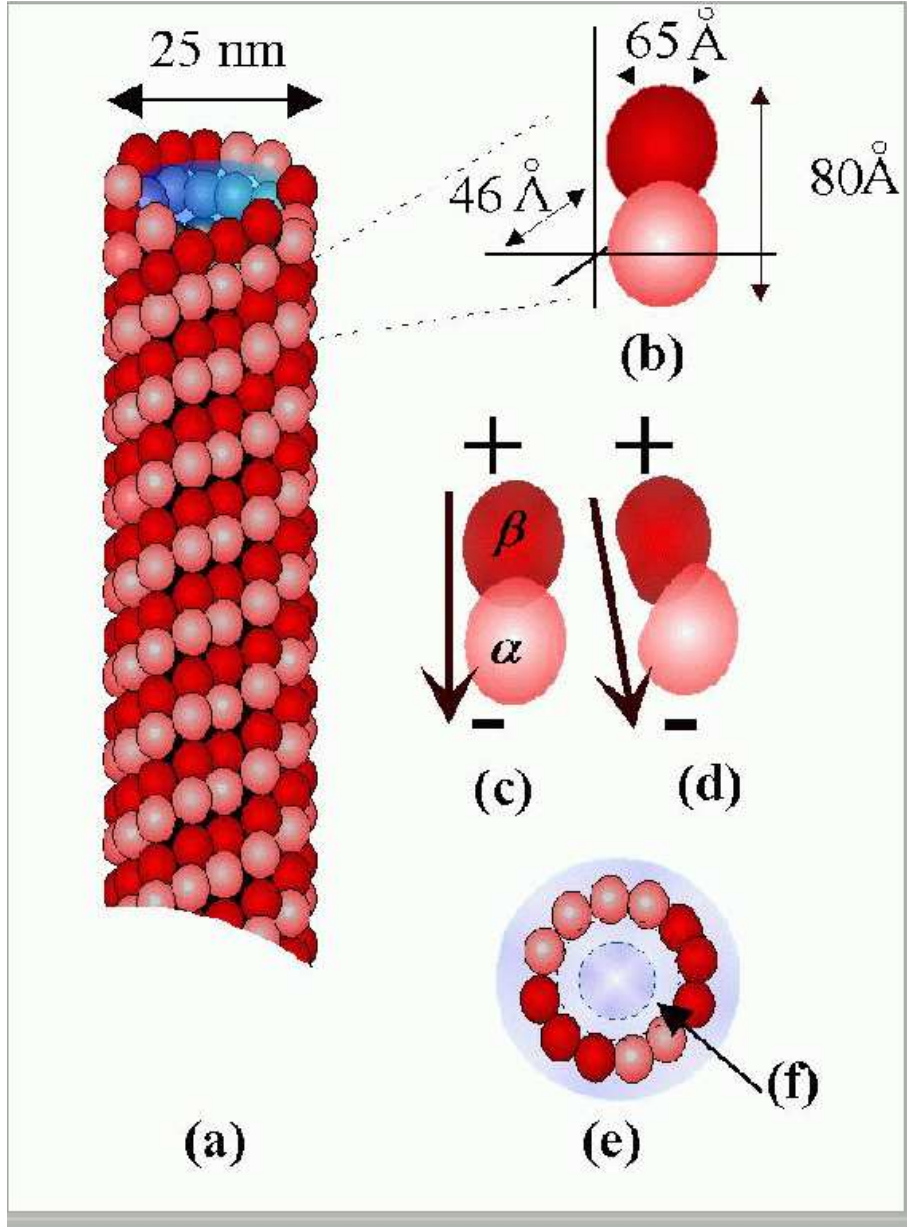}} \caption{{\it (a) Typical microtubule made of 13
tubulin protofilaments. Each protofilament is displaced vertically from its neighbor a fifth 
of the dimer vertical height. (b) dimensions of  the heterodimer as solved by electron 
crystallography of zinc-induced sheets ~\cite{nog} note that the MT consists of (c)
GTP-tubulin and (d) GDP-tubulin. Arrows indicate the direction of the electric dipole moment
for the two conformations. (e) a cross section of the MT showing the surrounding water
layer and (f) the thin interior layer that is to be treated as a QED cavity.}}
\end{centering}
\label{fig:FINAL_MT}
\end{figure}

Ferroelecricity is an essential ingredient of the
quantum-mechanical mechanism of energy transfer that we propose.
We have speculated~\cite{zioutas} that the ferroelectric nature of MTs, 
will be that of  {\it hydrated} ferroelectrics, i.e. the ordering of the electric  dipole moment of the tubulin molecules will be due to the interaction of the
tubulin dimers' electric dipoles with  the ordered-water molecules
in the interior and possibly exterior  of the microtubular
cavities. Ferroelectricity induces a dynamical dielectric
`constant' $\varepsilon (\omega)$  which is dependent on the
frequency $\omega$ of the excitations in the medium. Below a
certain frequency,  such materials are characterized by almost
vanishing dynamical dielectric `constants', which in turn implies
that electrostatic interactions inversely proportional to
$\varepsilon$ will be enhanced, and thus become dominant against
thermal losses. In the case of microtubules, the pertinent
interactions are of the electric dipole type, scaling with the
distance $r$ as $1/(\varepsilon r^3)$. For ordinary water media,
the dielectric constant is of order $80$. In the ferroelectric
regime however, this $\varepsilon$  is diminished significantly.
As a result, the electric dipole-dipole  interactions may overcome
the thermal losses proportional to $k_BT$ at room temperature inside the interior
cylindrical region of MT bounded by the dimer walls of thickness
of order of a few Angstroms~\cite{mn}. The situation is depicted
in Figure~\ref{fig:FINAL_MT}.

It is known experimentally~\cite{sackett}, that in a thin exterior neighborhood
of MTs there are areas of atomic thickness,
consisting of charged ions,  which isolate the MT from thermal
losses. This means that the  electrostatic interactions overcome
thermal agitations. It seems theoretically plausible, albeit yet unverified,
that such thermally isolated  exterior areas can also operate as
{\it cavity regions}, in a  manner similar to the areas interior
to MTs. At this point it is unclear whether there exist the
necessary coherent dipole quanta  in the ionic areas. Further
experimental and theoretical (simulational) work needs to be done
regarding this  issue and this is in progress.

Once such an isolation is provided, one can treat the thin
interior regions as electromagnetic cavities in a way similar to
that of QED cavities~\footnote{Note that the role of MT as {\it waveguides}
has been proposed by S. Hameroff already some time
ago~\cite{guides}. In our scenario on the other hand, we are
interested in isolated regions inside the MT which play the role
of {\it QED cavities}.}. QED cavities are well known for their
capacity to sustain in their interior coherent modes of
electromagnetic radiation. Similarly, one expects that such
coherent cavity modes will occur in the thin interior regions of
MTs bounded by the protein dimer walls. Indeed, as we discussed in
\cite{mn}, these modes are provided by the interaction of the
electric dipole moments of the ordered-water molecules in the
interior of MT with the quantised electromagnetic
radiation~\cite{delgiud,prep}. Such coherent modes are termed {\it
dipole quanta}. It is the interaction of such cavity modes with
the electric dipole excitations of the dimers that leads to the
formation of coherent (dipole) states on the tubulin dimer walls
of MTs. A review of how this can happen, and what {\it purely quantum} 
effects can emerge from the QED nature of MTs, will be the main topic of
this communication. 

In Section 2 we present a
concise expos\'{e} of our mechanism~\cite{mn} that justifies the
application of quantum physics to the treatment of certain aspects
of MT dynamics. In section 3 we develop an analogy of our
mechanism to the experimental setup used by Julsgaard et. al. In
Section 4 we present a straight forward calculation of how quantum
teleportation of states can occur in MTs, in direct analogy to the
suggester experimental quantum teleportation in optical cavities
that has been observed recently ~\cite{sz,hui}. We also draw a
parallel between certain geometrical features of MTs such as their
ordered structure which obeys a potentially information-encoding
code and suggest how this can be exploited for (quantum)
error-correction and dense coding. Section 5 contains our
conclusions and outlook.

\section{Quantum Coherence in Biological Matter?}

\subsection{Tubulin, Microtubules and Coherent States}

Tubulin is a common polar protein found mainly in the cytoskeleton of eukariotic cells which is especially enriched in brain tissue. Many of its properties have been studied both experimentally and theoretically because of its importance in mitosis, its role as the building block of microtubules and its relevance to several diseases including Cancer and Alzheimer's. Measurements ~\cite{nog} have confirmed earlier data suggesting that the tubulin heterodimer has dimensions 46 X 80 X 65 Angstrom [Fig. 1 (b)]. Under normal physiological conditions, tubulin is found as a heterodimer, consisting of two nearly identical monomers called the $\alpha-$ and $\beta-$ tubulin each of molecular weight of about 55kDalton ~\cite{nog} . MTs are hollow (25nm-outer diameter, 14nm inner diameter) -see Fig. 1 (a) tubes forming the main component of the cytoskeleton and apart from giving shape and support to the cell, they also play a major role in cellular transport and have been hypothesized to be central in cellular information processing. The interior of the MT, seems to contain {\it ordered water} molecules~\cite{mn} , which implies the existence of an
electric dipole moment and an electric field. We stress that the intracellular {\it ordered water} which is full of proteins and other molecules is different from ordinary water in various respects e.g. as is implied in ~\cite{water}.
It is to be understood that unless otherwise specified, from now on by 'tubulin' we refer to the $\alpha\beta$-dimer. Free tubulin can self-assemble into MTs both {\it in vivo} and {\it in vitro} where the most common arrangement of the tubulin dimers is such that, if one treats them as points, they resemble triangular lattices on the MT surface. The $\beta$-tubulin monomer of the heterodimer can bind guanosine 5' triphosphate (GTP) in which case it exists in an energy-rich form that favors polymerization, or it can bind guanosine 5' diphosphate (GDP-tubulin) thus being in an energy-poor form (GDP-tubulin) that favors dissociation. The structure of MTs has been the subject of comprehensive study and it transpires that MTs come in a variety of arrangements the predominant of which is a 5-start, period-13 helical tube of dimers which resembles a corn ear [Fig.1 (a)] made out of 13 offset protofilaments. Certain interesting phenomena arise during the (de-) polymerization of tubulin such as length oscillations, treadmilling etc. generally referred to as 'dynamic instability' and these have been studied extensively ~\cite{mtosci, mtstoch} but are not directly relevant to our analysis at this stage as such phenomena are suppressed in the remarkably stable axonal neural cytoskeleton. It has been shown that the GDP-GTP exchange (hydrolysis) releases approximately 0.42eV per molecule and is accompanied by a conformational change ~\cite{confchange}. This change has been modelled as resulting in a $27^{o}42'$ angle ~\cite{melki}  between the original line connecting the centers of the $\alpha$ and $\beta$ monomers and the new center-to-center line [Fig. 1 (d)]. As a result of this change in the geometry of the tubulin molecule, the orientation of the electric dipole moment also changes magnitude and direction. It has been put forward that each dimer has two hydrophobic protein pockets, containing $2 \times 18$ unpaired electrons ~\cite{dustin} that have at least two possible configurations associated with the GTP and GDP states of tubulin, which we will call $\uparrow$ and $\downarrow$ electron (or equivalently electric dipole moment) conformations respectively. 

Using the fact that a typical 'distance' for the transition between the
$\uparrow$ and $\downarrow$ conformations is of order of the distance between
the two hydrophobic dimer pockets, i.e. ${\cal O}(4~{\rm nm})$,
a simplistic estimate of the free tubulin electric dipole moment $d$ can be obtained based on a mobile charge of 36 electrons multiplied by this separation of 4nm giving a magnitude of $d = 2.3\times10^{-26} C\cdot m$ (or ~700 Debye) while using a more sophisticated molecular simulation, $d$ has been quoted at 1714 Debye ~\cite{brown}. It is evident that an experimentally determined electric dipole moment for the tubulin molecule and its dynamics are important areas of study that have to be undertaken if these studies are to move forward.

If we account for the effect of the water environment that screens the electric
charge of the dimers by the relative dielectric constant of the water, which 
is $\varepsilon/\varepsilon_0 \sim 80$, we arrive at a value of 
\be d_{dimer} \sim  3 \times 10^{-28}~ {\rm C} \times
{\rm m} \label{dipoledimer} \ee 
Note that under physiological conditions, 
the unpaired electric charges in the dimer may lead to even 
further suppression of $d_{dimer}$ (\ref{dipoledimer}).

At physiological pH (=7.2) MTs are negatively charged ~\cite{stebbins} due to the presence of a 15-residue carboxyl-terminus 'tail' and there have been suggestions that this C-terminus is important in polymerization, protein interactions and perhaps charge conduction ~\cite{sackett}. This terminus has not been included in the electron crystallography data of Nogales and Downing ~\cite{nog} so all values concerning the dipole moment are quoted with the understanding that $d$ has been calculated ignoring the effect of the C-terminus. It is also known that at pH 5.6 MTs become neutral. Finally, there have been some preliminary experiments aimed at measuring the electric field around MTs ~\cite{pokorny1,pokorny2,pokorny3} indicating that MTs could be ferroelectric, as we have suggested in our model of ~\cite{mn} and ~\cite{zioutas}. Note that although the 'caps' of the MT contain both GTP and GDP tubulin, it is well known experimentally ~\cite{melki} that the tubulin comprising the 'trunk' of the MT is GDP-tubulin incapable of acquiring a phosphate and becoming GTP tubulin.  However, this does not preclude electric-dipole moment flip wave propagation down the MT, as a flip at the cap can be propagated without phosphorylation or hyrdolysis but rather via the mechanism suggested below. In view of this, the value of the yet undetermined electric dipole moment direction flip angle $\theta _{flip}$ is much smaller than the $27^{o}42'$ value for free tubulin. The $\uparrow$ and $\downarrow$ states still exist though, but are hard to observe experimentally as they are not associated with a large-scale geometrical mass shift. Note that virtually all of the alternative suggested MT-based ``quantum brain" hypotheses today, fail to take this into account and instead wrongly suggest that $\theta_{flip}$ is of the order of $27^{o}$ and that such large distortions occur in the trunk of the polymerized MT. 

In standard models for the simulation of  MT dynamics~\cite{mtmodel}, the physical degree of freedom which is relevant for the description of energy transfer  is the
projection of the electric dipole moment on the longitudinal
symmetry axis (x-axis) of the MT cylinder. The $\theta _{flip}$ distortion
of the $\downarrow$-conformation leads to a displacement $u_n$ along
the $x$-axis. This way, the effective system is one-dimensional
(spatial), and one has the possibility of being quantum
integrable~\cite{mn}.

It has been suggested for quite some time that information
processing  via interactions among the MT protofilament chains can
be sustained on such a system, if the system is considered as a series
of interacting Ising chains on a triangular lattice. For such
schemes to work, one must first show that the electromagnetic
interactions  among the tubulin dimers are strong enough to
overcome thermal noise.  It is due to this problem that such
models for intra-neuronal information processing  have been
critisized as unphysical ~\cite{tegmark}. We shall return to this
issue later.  Classically, the various dimers can only be
in the $\uparrow$ and $\downarrow$ conformations. Each dimer is influenced
by the neighboring dimers resulting in the possibility of a
transition. This is the basis for classical information
processing, which constitutes the picture of a (classical)
cellular automaton.

If we assume (and there is good theoretical basis for such an assumption ~\cite{mn}) that each dimer can in fact find itself in a superposition of $\uparrow$ and $\downarrow$ states a {\it quantum nature} results. Tubulin can then be viewed as a typical {\it two-state}
quantum mechanical system, where the dimers couple to
conformational changes with $10^{-9}-10^{-11} {\rm sec}$
transitions, corresponding to an angular frequency $     \omega
\sim{\cal O}( 10^{10}) -{\cal O}(10^{12})~{\rm Hz}$. In the
present work we assume the upper bound of this frequency range to
represent (in order of magnitude) the characteristic frequency of
the dimers, viewed as a two-state quantum-mechanical system: \be
     \omega _0 \sim {\cal O}(10^{12})~{\rm Hz}
\label{frequency2} \ee 

As we shall see below, such a frequency range
is not unusual in biology.

Let $u_n$ be the displacement field of the $n$-th dimer in a MT
chain. The continuous approximation proves sufficient for the
study of phenomena associated with energy transfer in biological
cells, and this implies that one can make the replacement \be
  u_n \rightarrow u(x,t)
\label{three} \ee with $x$ a spatial coordinate along the
longitudinal symmetry axis of the MT. There is a time variable $t$
due to fluctuations of the displacements $u(x)$ as a result of the
dipole oscillations in the dimers.

The effects of the neighboring dimers (including neighboring
chains) can be phenomenologically accounted for by an effective
potential $V(u)$. In the model of ref. \cite{mtmodel} a
double-well potential was used, leading to a classical kink
solution for the $u(x,t)$ field. More complicated interactions are
allowed in the picture of ref. \cite{mn}, where we have considered more generic
polynomial potentials.

The effects of the surrounding water molecules can be accounted for
by a viscous force term that damps out the dimer oscillations, \be
 F=-\gamma \partial _t u
\label{six} \ee with $\gamma$ determined phenomenologically at
this stage. This friction should be viewed as an environmental
effect, which however does not lead to energy dissipation, as a
result of the non-trivial solitonic structure of the ground-state
and the non-zero constant force due to the electric field. This is
a well known result, directly relevant to energy transfer in
biological systems \cite{lal}.

The effective equation of motion for the
relevant field degree of freedom $u(x,t)$ reads: \be u''(\xi) +
\rho u'(\xi) = P(u) \label{generalsol} \ee where $\xi=x-vt$, $v$
is the velocity of the soliton, $\rho \propto
\gamma$~\cite{mtmodel}, and $P(u)$ is a polynomial in $u$, of a
certain degree, stemming from the variations of the potential
$V(u)$ describing interactions among the MT chains~\cite{mn}. In
the mathematical literature~\cite{otinowski} there has been a
classification of solutions of equations of this form. For certain
forms of the potential~\cite{mn} the solutions include {\it  kink
solitons} that may be responsible for dissipation-free energy
transfer in biological cells~\cite{lal}:
\begin{equation}
u(x,t) \sim c_1 \left(tanh[c_2(x-v t)] + c_3 \right) \label{kink}
\end{equation}
where $c_1,c_2, c_3$ are constants depending on the parameters of
the dimer lattice model. For the form of the potential assumed in
the model of \cite{mtmodel} there are solitons of the form $
u(x,t)= c_1' + \frac{c_2' - c_1'}{1 + e^{c_3'(c_2'-c_1')(x -
vt)}}$, where again $c_i', i=1,\dots 3$ are appropriate constants.

A semiclassical quantization of such solitonic states has been
considered in \cite{mn}. The result of such a quantization yields
a modified soliton equation for the (quantum corrected) field
$u_{q}(x,t)$ \cite{tdva} \be
    \partial ^2 _t u_q(x.t) - \partial _x ^2 u_q(x,t)
+ {\cal M}^{(1)} [u_q(x,t)] = 0 \label{22c} \ee
with the notation \\
$M^{(n)} = e^{\frac{1}{2}(G(x,x,t)-G_0(x,x))\frac{\partial ^2}
{\partial z^2}} U^{(n)}(z) |_{z=u_q(x,t)}$, and $U^{(n)} \equiv
d^n U/d z^n$. The quantity $U $ denotes the potential of the
original soliton Hamiltonian, and $G (x,y,t)$ is a bilocal field
that describes quantum corrections due to the modified boson field
around the soliton. The quantities $M^{(n)}$ carry information
about the quantum corrections. For the kink soliton (\ref{kink})
the quantum corrections (\ref{22c}) have been calculated
explicitly in ref. \cite{tdva}, thereby providing us with a
concrete example of a large-scale quantum coherent state.

A typical propagation velocity of the kink solitons (e.g. in the
model of ref. \cite{mtmodel}) is $v \sim 2~{\rm m/sec}$, although,
models with $v \sim 20~m/sec$ have also been
considered~\cite{satar}. This implies that, for moderately long
microtubules of length $L \sim 10^{-6}$ m, such kinks transport
energy without dissipation in \be
      t_F \sim 5 \times 10^{-7}~{\rm sec}
\label{FS} \ee 
Energy will be transferred super-efficiently via this mechanism only if the decoherence time is of the order of, or longer than, this time. We shall see in fact that indeed such time
scales are comparable to, or smaller in magnitude than, the
decoherence time scale of the coherent
(solitonic) states $u_q (x,t)$. This then implies that
fundamental quantum mechanical phenomena may be responsible
for frictionless, dissipantionless super-efficient energy (and signal) transfer and/or transduction across microtubular networks in the cell.

\subsection{Microtubules as Cavities}

In ref. \cite{mn} we have presented a microscopic analysis of the
physics underlying the interaction of the water molecules with the
dimers of the MT, which is responsible for providing the friction
term (\ref{six}) in the effective (continuum) description. Below
we briefly review this scenario.

As a result of the ordered
structure of the water environment in the interior of MTs, there
appear {\it collective} coherent modes, the so-called dipole
quanta~\cite{delgiud}. These arise from the interaction of the
electric dipole moment of the water molecule with the quantized
radiation of the electromagnetic field~\cite{prep}, which may be
self-generated in the case of MT arrangements~\cite{satar,mn}.
Such coherent modes play the role of `cavity modes' in the quantum
optics terminology. These in turn interact with the dimer
structures, mainly through the unpaired electrons of the dimers,
leading to the formation of a quantum coherent solitonic state
that may extend even over the entire MT network. As mentioned
above, such states may be identified~\cite{mn} with semi-classical
solutions of the friction equations (\ref{generalsol}). These
coherent, almost classical, states should be viewed as the result
of {\it decoherence} of the dimer system due to its
interaction/coupling with the water environment~\cite{zurek}.

Such a dimer/water coupling can lead to a situation analogous to
that of atoms interacting with coherent modes of the
electromagnetic radiation in {\it quantum optical cavities}, namely
to the so-called {\it Vacuum-Field Rabi Splitting} (VFRS)
effect~\cite{rabi}. VFRS appears in both the emission and
absorption spectra of atoms~\cite{agar} in interaction with a
coherent mode of electromagnetic radiation in a cavity. For our
purposes below, we shall review the phenomenon by restricting
ourselves for definiteness to the absorption spectra case.

Consider a collection of $N$ atoms of characteristic 
frequency $\omega_0$ inside an electromagnetic cavity. Injecting a pulse of 
of frequency $\Omega$ into the cavity causes a doublet
structure (splitting) in the absorption spectrum of the
atom-cavity system with peaks at: \be \Omega = \omega _0 -
\Delta/2 \pm \frac{1}{2}( \Delta ^2 + 4 N \lambda ^2 )^{1/2}
\label{rabiabs} \ee where $\Delta = \omega_c - \omega_0 $ is the
detuning of the cavity mode, of frequency $\omega_c$, compared to
the atomic frequency. For resonant cavities the splitting occurs
with equal weights \be
  \Omega = \omega_0 \pm \lambda \sqrt{N}
\label{rabisplitting} \ee Notice here the {\it enhancement} of the
effect for multi-atom systems $N >> 1$. The quantity  $2\lambda
\sqrt{N}$ is called the `Rabi frequency'~\cite{rabi}. {}From the
emission-spectrum analysis an estimate of $\lambda$
can be inferred which involves the matrix element, ${\underline
d}$, of atomic electric dipole between the energy states of the
two-level atom~\cite{rabi}: \be
   \lambda = \frac{E_{c}{\underline d}.{\underline \epsilon}}{\hbar}
\label{dipolerabi} \ee where ${\underline \epsilon}$ is the cavity
(radiation)  mode polarisation, and \be E_{c} \sim
\left(\frac{2\pi \hbar \omega_c}{\varepsilon  V}\right)^{1/2}
\label{amplitude} \ee is the r.m.s. vacuum (electric) field
amplitude at the center of a cavity of volume $V$, and of
frequency $\omega_c$, with $\varepsilon $ the dielectric constant
of the medium inside the volume $V$. In atomic physics the VFRS
effect has been confirmed by experiments involving beams of
Rydberg atoms resonantly coupled to superconducting
cavities~\cite{rabiexp}.

In the analogy between the thin cavity regions near the dimer
walls of MTs with electromagnetic cavities, the
role of atoms in this case is played by the unpaired two-state
electrons of the tubulin dimers~\cite{mn} oscillating with a
frequency (\ref{frequency2}). To estimate the Rabi coupling
between cavity modes and dimer oscillations, one should use
(\ref{dipolerabi}) for the MT case. 

We have used some simplified models for the ordered-water
molecules, which yield a frequency of the coherent
dipole quanta (`cavity' modes) of order~\cite{mn}: \be
     \omega_c \sim 6 \times 10^{12} s^{-1}
\label{frequency} \ee Notably this is of the same order of magnitude as the
characteristic frequency of the dimers (\ref{frequency2}),
implying that the dominant cavity mode and the dimer system are
almost in resonance in our model of \cite{mn}. Note that this
is a feature shared by atomic physics systems in cavities, and
thus we can apply the pertinent formalism to our system. Assuming
a relative dielectric constant of water w.r.t to that of vacuum
$\epsilon_0$, $\varepsilon/\varepsilon_0 \sim 80$, one obtains
from (\ref{amplitude}) for the case of MT cavities: \be
    E_{c} \sim 10^{4}~{\rm V/m}
\label{eowmt} \ee Electric fields of such a magnitude can
be provided by the electromagnetic interactions of the MT dimer
chains, the latter viewed as giant electric
dipoles~\cite{mtmodel}. This suggests that the
coherent modes $\omega_c$, which in our scenario interact with the
unpaired electric charges of the dimers and produce the kink
solitons along the chains, owe their existence to  the (quantized)
electromagnetic interactions of the dimers themselves.

The Rabi coupling for the MT case then is estimated from
(\ref{dipolerabi}) to be of order: \bea &~&    {\rm
Rabi~coupling~for~MT} \equiv \lambda _{MT}
= \nonumber \\
&~& \sqrt{{\cal N}} \lambda_0 \sim 3 \times 10^{11} s^{-1}
\label{rabiMT} \eea which is, on average, an order of magnitude
smaller than the characteristic frequency of the dimers
(\ref{frequency2}).

In the above analysis, we have assumed that the system of tubulin dimers
interacts with a {\it single} dipole-quantum coherent mode of the
ordered water and hence we ignored dimer-dimer interactions. 
More complicated cases, involving interactions
either among the dimers or of the dimers with more than one
radiation quantum, which undoubtedly occur {\it in vivo}, may affect
the above estimate.

The presence of such a coupling between water molecules and dimers
leads to quantum coherent solitonic states of the electric dipole
quanta on the tubulin dimer walls. To estimate the decoherence
time we remark that the main source of dissipation (environmental
entanglement) comes from the imperfect walls of the cavities,
which allow leakage of coherent modes and energy. 
The time scale, $T_r$, over
which a cavity-MT dissipates its energy, can be identified in our
model with the average life-time $t_L$ of a coherent-dipole
quantum state, which has been found to be~\cite{mn}: $T_r \sim t_L
\sim  10^{-4}~{\rm sec}$. This leads to a first-order-approximation estimate of the
quality factor for the MT cavities, $Q_{MT} \sim \omega_c T_r \sim
{\cal O}(10^8)$. We note, for comparison, that high-quality
cavities encountered in Rydberg atom experiments dissipate energy
in time scales of ${\cal O}(10^{-3})-{\cal O}(10^{-4})$ sec, and
have $Q$'s which are comparable to $Q_{MT}$ above. The analysis of
~\cite{mn} then yields the following estimate for the collapse
time of the kink coherent state of the MT dimers due to
dissipation: \be t_{collapse} \sim {\cal O}(10^{-7})-{\cal
O}(10^{-6})~{\rm sec} \label{tdecohsoliton} \ee This is larger
than the time scale (\ref{FS}) required for energy transport
across the MT by an average kink soliton in the models of
\cite{mtmodel,satar}. The result (\ref{tdecohsoliton}), then,
implies that quantum physics is relevant as far as
dissipationless energy transfer across the MT is concerned.

In view of this specific model, we are therefore in stark \textit{disagreement}
with the conclusions of Tegmark in \cite{tegmark}, i.e. that only
classical physics is relevant for studying the energy and signal transfer in
biological matter. Tegmark's conclusions did not take proper account of the
possible isolation against environmental interactions, which seems to occur
inside certain regions of MTs with 
appropriate geometry and properties MT.

We would now like to discuss the feasibility of the above, admittedly speculative, ideas
by making a brief report on recent progress made by 
experimentally demonstrating macroscopic quantum entanglement
at room temperature in atomic physics.

\subsection{On Ordered Water in Biological Systems}

The above scenaria may find some support by independent
studies of water in biological matter, which we summarize below.
Recent experimental spectroscopic studies of
resonant intermolecular  transfer of vibrational energy in liquid
water~\cite{woutersen} have established that  energy is transfered
extremely rapidly and along many water molecules before it
dissipates.  This energy is in the form of OH-stretch excitations
and is thought to be mediated by  dipole-dipole interactions in
addition to a yet unknown mechanism  which speeds up the transfer
beyond that predicted by the so-called  F\"orster expression for
the energy trasfer rate between two OH oscillators, k.
\be k=
T_1^{-1}\left(\frac{r_o}{r}\right)^{6}  \label{forster}
\ee
where
$T_1$ is the lifetime of the excited state, $r$ the distance
between the oscillators and $r_o$  the F\"orster radius. The
F\"orster radius, which is a parameter experimentally determined
for each material, characterizes the intermolecular energy
transfer and has been determined by Woutersen et.
al.~\cite{woutersen} to be $r_o = 2.1 \pm 0.05$~Angstroms  while
the typical intermolecular distance (at room temperature) for
water is $\sim 2.8$~Angstrom. It is evident from these data that
the energy transfer in pure water will be extremely fast  (of
order 100ps) and yet experimentally it is determined to be even
faster than that. Woutersen {\it et. al.} speculate that this extremely
high rate of  resonant energy transfer in liquid water may be a
consequence of the proximity of the  OH groups in liquid water
which causes other, higher-order -uples to also exchange energy.
We propose another mechanism to explain the rapidity of the energy
transfer, namely kink-soliton propagation. This proposition is
based on the phenomenological realization that this  is exactly
the kind of energy transfer that one would expect to see
experimentally as a result of the existence of kink-solitons.  It
is evident that such a mechanism, regardless of exact origin, is
{\it ideal} for loss-free energy  transfer between OH groups
located on either different biomolecules or along extended
biological structures such as MTs which would be covered (inside
and out) with water. Note also that such a mechanism would predict
that OH groups in hydrophobic environments would be able to remain
in a vibrationally excited state longer than OH groups in
hydrophilic environment lending credence to our working assumption
that the electrons inside the hydrophobic pockets of the tubulin
molecules are sufficiently isolated from thermal noise.\\

It must be stressed though that such solitonic states in water
may not be quantum in origin in the case of microtubules.
The 25 nm diameter of the MT is too big a region to allow for quantum
effects to be sustained throughout, as we discussed above.
Such solitons may be nothing other than the ones conjectured in
\cite{jibu}, which may be responsible for the optical
transparency of the water interior of MTs.
However, such classical solitons in the bulk of the
water interior may co-exist with the quantum coherent states on the dimer
walls~\cite{mn}.

\subsection{Error Correction and Long-lived Quantum
Entanglement of Macroscopic Sample at Room Temperature}

As we have seen above, under appropriate environmental isolation,
it is possible to obtain quantum coherence on {\it macroscopic
populations} of tubulin dimers in microtubule systems,
which can be sustained for long enough times so that dissipationless
energy and signal (information) transfer can occur in a cell.

In a recent article~\cite{caes} Julsgaard et. al. describe the
macroscopic  entanglement of two samples of Cs atoms at room
$T^o$. The entangling mechanism is a pulsed laser beam and
although the atoms are  far from cold or isolated from the
environment,  partial entanglement of bulk spin is unambiguously demonstrated
for  $10^{12}$ atoms for $\sim 0.5 ms$.  The system's resilience
to decoherence is in fact {\it facilitated} by the existence of a large
number of atoms as even though atoms lose the proper  spin
orientation continuously, the {\it bulk} entanglement is not
immediately lost.  Quantum informatics, the science that deals
with  ways to encode, store and retrieve information written in
qubits has to offer an alternative  way of interpreting the
surprising resilience of the Cs atoms by using the idea of ``redundancy".
Simply  stated, information can be stored in such a way that the
logical (qu)bits correspond to many physical  (qu)bits and thus
are resistant to corruption of content. Yet another way of looking
at this is given in the work by Kielpinski et.
al.~\cite{kielpinski} where they have experimentaly demonstrated a
decoherence-free quantum memory of one qubit by encoding the qubit
into the ``decoherence-free subspace" (DFS) of a pair of trapped Berrilium 
$^9Be^+$ ions. They achieved this by exploiting a
"safe-from-noise-area" of the Hilbert space for a {\it
superposition}  of two basis states for the ions, thus encoding
the qubit in the superposition rather  than one of the basis
states. By doing this they achieved decoherence times on average  an
order of magnitude longer. \\ Both of the above works show that it
is possible to use DFS, error correction and high  redundancy to
both store information and to keep superpositions and
entanglements alive for biologically relevant times in macroscopic systems at high
temperature.\\
Thus it nay  not be entirely inappropriate to
imagine that in biological {\it in vivo} regimes,
one has, under certain circumstances, such as 
specified above, similar entanglement of tubulin/MT arrangements.

\section{Possible Implications
of Quantum Coherence to the Functioning of Cells}

The above raises the question of how such phenomena can affect the
functioning of cells. In other words, would the existence of such
coherent states and the emergence of quantum mechanical
entanglement be somehow useful or beneficial to biological function? 
Is it then reasonable to propose that in certain cases, natural selection may have favored molecules and cellular structures that 
exhibited such phenomena? If we accept the
notion that according to the laws of quantum physics certain macroscopic
arrangements of atoms will exhibit such effects, is it not reasonable
then to expect that biomolecules and (by extension) cellular structures
and whole cells have 'found' a use for such phenomena and have
evolved to incorporate them? We stress that at a given instant in time, the different microtubule coherent states participating in a specific bulk entanglement would be almost identical due to the fact that they are related/triggered by a specific ``external agent" (e.g. the passing of a specific train of action potentials.) This is of outmost importance since it increases the system's resilience to decoherence (by entangling a large number of nearly identical states), in addition to facilitating "sharp decision making" (i.e. rapid choice among a vast number of very similar states) as explained in ~\cite{brazil} which is presumably a trait favored by natural selection.
Here we digress to investigate one possible use of such effects by noting a straight-forward
application of entanglement to {\it teleportation} of coherent
quantum states across and between cells.

We define teleportation as the complete transfer of the {\it
coherent state} of an MT {\it without any direct transfer of mass
or energy}. This means that the 'receiver' MT finds itself in an identical
state to the 'sender' MT.
We will demonstrate that given the possibility for
entangled states, teleportation between microtubule A and
microtubule C can happen as follows:

A coherent state in microtubule A (referred to as simply A and
designated as $\ket{\Psi(A)} $ )  of the (collective) dipole
moment(s) being in either of the two classically allowable states
with probability amplitude $\omega_{0}$ and $\omega_{1}$ can be
written as:

\be \ket{\Psi(A)}=\omega_{0}\ket{0}+\omega_{1}\ket{1} \ee

Step 1: The cell finds itself with {\it microtubule B and
microtubule C} -which can be close together or collinear- in an
entangled state written as:

\be
\ket{\Psi(B,C)}=\frac{1}{\sqrt{2}}\biggl(\ket{1_B,0_C}+\ket{0_B,1_C}\biggr)
\ee

The combined state of A,B,C can be written as:

\be \ket{\Psi(A,B,C)}=\ket{\Psi(A)} \otimes \ket{\Psi(B,C)} \ee

which upon expanding the outer product can be written as:

\be
\ket{\Psi(A,B,C)}=\frac{1}{\sqrt{2}}\biggl(\omega_{0}(\ket{0_{A},1_{B},0_{C}}+
\ket{0_{A},0_{B},1_{C}})+\omega_{1}(\ket{1_{A},1_{B},0_{C}}+\ket{1_{A},0_{B},1_{C}})\biggr)
\ee

We can also express the combined state $\ket{\Psi(A,B,C)}$ in a
different basis, known as the ``Bell basis". Instead of $\ket{0}$
and $\ket{1}$, the basis vectors will now be,

\be
\ket{\Psi^{\pm}(A,B)}=\frac{1}{\sqrt{2}}\biggl(\ket{0_{A},1_{B}}\pm\ket{1_{A},0_{B}}\biggr)
\ee and \be
\ket{\Phi^{\pm}(A,B)}=\frac{1}{\sqrt{2}}\biggl(\ket{0_{A},0_{B}}\pm\ket{1_{A},1_{B}}\biggr)
\ee

In this new basis, our state of the three microtubules
$\ket{\Psi(A,B,C)}$ is written as:

\begin{eqnarray}
&~& \ket{\Psi(A,B,C)}=\frac{1}{2}\biggl(\ket{\Psi^{+}(A,B)}\otimes(\omega_{0}\ket{0_{C}}+
\omega_{1}\ket{1_{C}})+ \nonumber \\
&~& \ket{\Phi^{+}(A,B)}\otimes(\omega_{0}\ket{1_{C}}+\omega_{1}\ket{0_{C}})+
\ket{\Psi^{-}(A,B)}\otimes(\omega_{0}\ket{0_{C}}-\omega_{1}\ket{1_{C}})+
\nonumber \\
&~& \ket{\Phi^{-}(A,B)}\otimes(\omega_{0}\ket{1_{C}}-\omega_{1}\ket{0_{C}})\biggr)
\end{eqnarray}

This concludes the first step of teleporting the state of MT A to
MT C.\\

Step 2:

Notice that so far, MT A has not been manipulated by the cell,
i.e. the coherent state of A which we designated as
$\ket{\Psi(A)}=\omega_{0}\ket{0}+\omega_{1}\ket{1} $ has not been
touched. Now the part of the cell containing A and B (let's call
it the ``sender part") makes a "measurement" -which in our case can
be an electromagnetic interaction with a passing action potential
or the binding of a MAP molecule. If this ``measurement" or forced
collapse is done in the Bell basis, on $\ket{\Psi^{\pm}(A,B)}$ it
will project the state in MT C (!) to:

\be \ket{\Psi^{\pm}(C)}=\langle\Psi^{\pm}(A,B)|\Psi(A,B,C)\rangle=
\omega_{0}\ket{0_{C}}\pm\omega_{1}\ket{1_{C}} \ee

similarly

\be
\ket{\Phi^{\pm}(A,B)}\longrightarrow\ket{\Phi^{\pm}(C)}=\omega_{0}\ket{1_{C}}\pm\omega_{1}\ket{0_{C}}
\ee

This effectively concludes the teleportation of the state of MT A
to MT C with one caveat. There is a probabilistic nature to this process,
which means that MT C may receive the exact copy of the state of
MT A i.e. $\ket{\Psi^{+}(C)}$ {\it or} it may receive a state
which is a unitary transformation away from the original
$\ket{\Psi(A)}$ (one of the other three possibilities:
$\ket{\Psi^{-}}$ or $\ket{\Phi^{\pm}}$. MT C can reproduce the
state of MT A if there is a 'hardwired' condition that when MT C
receives $\ket{\Psi^{+}}$ it does nothing further, yet if it
receives one of the other three, it performs the correct unitary
transformation to obtain the correct state from A. This
'hardwired' behavior can be implemented through the use of codes,
not unlike the Koruga bioinformation \cite{kor} code that MTs
follow.

Teleportation is a direct consequence of the existence of
entanglement and can be imagined as the basis of intra- and inter-
cellular correlation which leads to yoked function (e.g.
intracellulary during translation and intercellulary during yoked
neuron firing). Experiments to check for such teleportation of
states can be designed based on the Surface Plasmon Resonance
(SPR) paradigm \cite{spr} as applied to sheets of polymerized
tubulin immobilized on a dextran-layered gold film.

Note that our use of pure state vectors, $|\Psi\rangle$,
to describe the coherent states along a MT arrangement is justifiable since
they do not obey the ordinary Schr\"odinger evolution equation.
Instead, they obey the stochastic equations of open systems,
of the form discussed in \cite{gisin}. Nowhere in our
proof of teleportation above did we make use of the presice form of the
evolution equations. As argued in \cite{gisin},
by using appropriate stochastic (Langevin type) equations
one may recover, for instance,
the standard Lindblad form of evolution equations
for the corresponding density matrices 
$\rho ={\rm Tr}_{\cal M}|\Psi\rangle\langle \Psi |$,
where ${\cal M}$ is an appropriate subset of environmental
degrees of freedom, non-accessible to the observer.

\begin{figure}[!ht]
\begin{centering}
\scalebox{0.6}{\epsfig{figure=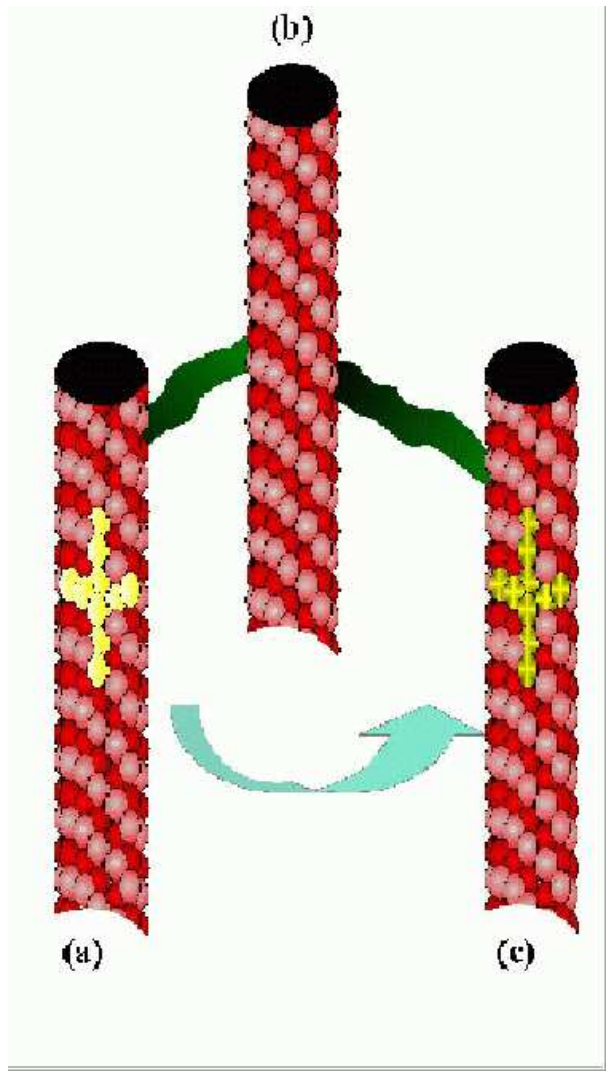}} \caption{{\it Schematic of a microtubular 
quantum teleportation of states. MT a sends its state (represented by a cross) to MT c 
without any transgfer of mass or energy. Both MT a and MT c are entangled with MT b (entanglement represented by presence of connecting MAPs.}}
\end{centering}
\label{fig:teleport}
\end{figure}

\section{Conclusions and Outlook}

In \cite{mn} we have put forward a comprehensive model conjecture treating certain regions inside MTs as {\it isolated} high-Q(uality) QED {\it cavities}. We
also presented a scenario according to which the presence of 
ordered water in the interior of MTs 
results in the appearance of electric dipole quantum coherent
modes, which couple to the unpaired electrons of the MT dimers via
Rabi vacuum field couplings. The situation is analogous to the
physics of Rydberg atoms in electromagnetic cavities~\cite{rabi}.
In quantum optics, such couplings may be considered as
experimental proof of the quantized nature of the electromagnetic
radiation. In our case, therefore, if experimentally detected, such couplings
would indicate the existence of coherent quantum modes of
electric dipole quanta in the ordered water environment of MT,
as conjectured in ref. \cite{delgiud,prep}, and used here.

To experimentally verify such a situation, one must first try to detect the emergent ferroelectric properties of MTs, which are predicted by this model and are potentially observable. Measurement of the dipole moment of the tubulin dimers is also an important step. A suggestion along these lines has been put forward in ref. \cite{zioutas}.

In addition, one should verify the aforementioned vacuum field
Rabi coupling (VFRS), $\lambda_{MT}$, between the MT dimers and
the ordered water quantum coherent modes. The existence of this coupling, could be tested experimentally by the same methods used
to measure VFRS in atomic physics~\cite{rabiexp}, i.e. by using
the MTs themselves as {\it cavity environments}, in the way
described above, and considering tunable probes to excite the
coupled dimer-water system. Such probes could be pulses of
(monochromatic) light coupling to MTs. This would be the analogue of an external
field in the atomic experiments mentioned above. The field would
then resonate, not at the bare frequencies of the coherent dipole
quanta or dimers, but at the {\it Rabi splitted} ones, leading to
a double peak in the absorption spectra of the
dimers~\cite{rabiexp}. By using MTs of different sizes one could
thus check on the characteristic $\sqrt{N}$-enhancement of the
(resonant) Rabi coupling (\ref{rabisplitting}) for MT systems with
$N$ dimers.

In the quantum-mechanical scenario for MT dynamics discussed above, as suggested in \cite{mn}, a quantum-hologram picture for information processing of
MT networks emerges. Further, the existence of solitonic quantum-coherent states along the MT dimer walls implies a role for these biological entities as
logic gates~\cite{mavrogates}. 
Consider, for instance, a node (junction) of
three MTs connected by microtubule associated proteins (MAPs) see Fig. 3. The quantum nature of the coherent states makes the junction interaction {\it probabilistic}.
Therefore at tube junctions one is facing a {\it Probabilistic
Boolean Interaction}~\footnote{We would like to thank Dr. Deeph Chana
for suggesting this terminology.}. The probability for having a
solitonic coherent state in a MT branch does depend on its
geometric characteristics (such as length ). By modulating the length of the tubes and the binding sites of the MAPs one a bias can be introduced 
between bit states which can affect the probabilistic final
outcomes. This has obvious implications for information processing
by MT networks.

\begin{figure}[!ht]
\begin{centering}
\scalebox{0.6}{\epsfig{figure=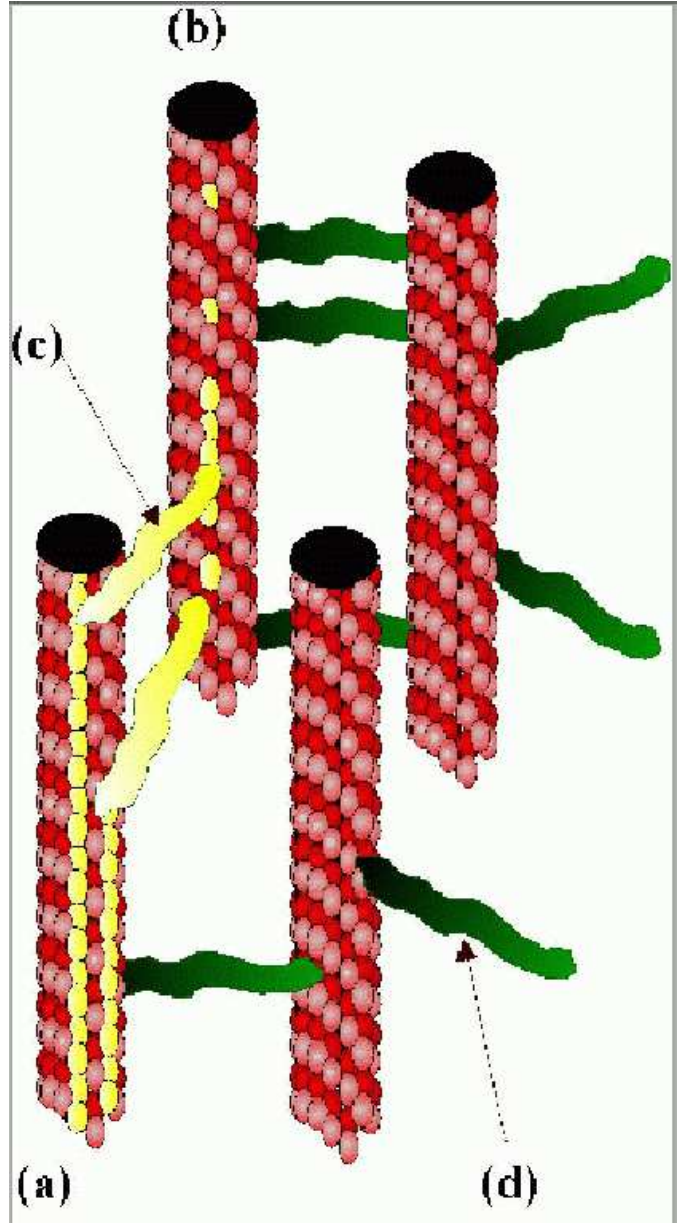}} \caption{{\it Example of a microtubular logic
gate: A XOR logic gate where ``0" is represented by absence of soliton and ``1" by presence of 
soliton.(a)Input MT. (b) Output MT. (c)A MAP transmitting a soliton. (d) A ``quiet" MAP. 
MT a has two solitons travelling, encountering two MAPs that transmit both  solitons to MT b (b).In this hypothetical scenario, the solitons arrive out of phase at MT b and cancel each 
other out.   The truth table for XOR reads: 
$0,0\rightarrow0;   
0,1\rightarrow1;  
1,0\rightarrow1, 
1,1\rightarrow0.$  
and in this case is realized by MTs if the MAPs are arranged such that each can transmit a soliton independently but if they both transmit, the solitons cancel out.}} \end{centering} \label{fig:XOR} \end{figure}

Such a binary information system can then provide the basic substrate for
quantum information processing inside a (not exclusively neural) cell. In a typical MT network, there may be about $10^{12}$
tubulin dimers. Such a number is macroscopic, and one is tempted
to express doubt as to whether, in realistic biological
situations, such macroscopic populations of `particles' can be
entangled quantum mechanically, with the entangled state being
maintained for a relatively long period of time. It
is worth stressing again that in atomic physics the
experiments of ref.~\cite{caes} have demonstrated {\it experimentally} the
existence of long-lived entangled states of {\it macroscopic}
populations of Cs gas samples, each sample containing
$10^{12}$ atoms. In such experiments entanglement is generated via
interaction with pules of light. Thus it is not impossible that {\it in vivo} one has, under certain circumstances as specified above, similar entanglement of MT coherent quantum states.

If it is experimentally confirmed that treating MTs as QED cavities is a fair approximation to their function, then it may not be so foolhardy to imagine that nature has provided us with the necesasary structures (microtubules) to operate as the basic substrate for {\it quantum computation}
either {\it in vivo}, e.g. in the way the brain works, or {\it in vitro},
i.e. it would allow us to construct quantum computers by using
microtubules as building blocks, in much the same way as QED cavities in quantum optics are currenlty being used in successful attempts at implementing qubits ~\cite{hui}.\\

\section*{Acknowledgements} 

We would like to thank Prof. S. Zubairy for useful discussions concerning quantum teleportation by QED cavities. 
We would also
like to thank 
Dr. P. Eagles, Prof. M. Holwill, Dr. M. Damzen, Prof. A. Michette,
Dr. K. Powell and Prof. Sarben Sarkar 
for discussions and their interest in further 
pursuing some of the experimental aspects of our work.

\end{document}